\documentclass[lang=en,11pt,a4paper]{elegantpaper}

\usepackage{graphicx} % Essential for including images
\graphicspath{ {./figures/} } % Set path for graphics files
\usepackage[table]{xcolor} % Color support (includes tables)
\usepackage{hyperref} % For clickable links and references - LOAD LATE
\usepackage{pdfpages}

\usepackage{bbding}
\usepackage{amsfonts,amsmath,amssymb,amsthm,mathtools} % Comprehensive math support

\usepackage{float} % Provides more float placement options (use [H] sparingly)
\usepackage{subcaption} % For multi-panel figures (subfigures)
\usepackage{authblk}
\title{Evolution without an Oracle: Driving Effective Evolution with LLM Judges}

\author{
    Zhe Zhao\textsuperscript{1}\thanks{\Envelope\ Correspondence to: \href{mailto:ehz0002@stanford.edu}{ehz0002@stanford.edu} and \href{mailto:zz8680@princeton.edu}{zz8680@princeton.edu}}, 
    Yuheng Yang\textsuperscript{2}, 
    Haibin Wen\textsuperscript{3}, 
    Xiaojie Qiu\textsuperscript{1}, 
    Zaixi Zhang\textsuperscript{4}\textsuperscript{*}, 
    and Qingfu Zhang\textsuperscript{3}
    \vspace{1em} \\ 
    \small\itshape
    \textsuperscript{1}Stanford University \quad
    \textsuperscript{2}Independent Researcher \\
    \textsuperscript{3}City University of Hong Kong \quad
    \textsuperscript{4}Princeton University
}

\usepackage{multirow} % For multi-row cells in tables
\usepackage{booktabs} % For professional quality table rules (\toprule, etc.)

\usepackage{algorithm} % Environment for algorithm descriptions
\usepackage{algpseudocode} % Pseudocode style for algorithms

\usepackage{listings} % For typesetting code snippets
\usepackage{array}

\begin{document}
\maketitle
\begin{abstract}
The integration of Large Language Models (LLMs) with Evolutionary Computation (EC) has unlocked new frontiers in scientific discovery but remains shackled by a fundamental constraint: the reliance on an \textbf{Oracle}—an objective, machine-computable fitness function. This paper breaks this barrier by asking: \textit{Can evolution thrive in a purely subjective landscape governed solely by LLM judges?} We introduce MADE (Multi-Agent Decomposed Evolution), a framework that tames the inherent noise of subjective evaluation through "Problem Specification." By decomposing vague instructions into specific, verifiable sub-requirements, MADE transforms high-variance LLM feedback into stable, precise selection pressure. The results are transformative: across complex benchmarks like DevAI and InfoBench, MADE outperforms strong baselines by over 50\% in software requirement satisfaction (39.9\% $\to$ 61.9\%) and achieves a 95\% perfect pass rate on complex instruction following. This work validates a fundamental paradigm shift: moving from optimizing "computable metrics" to "\textbf{describable qualities}," thereby unlocking evolutionary optimization for the vast open-ended domains where no ground truth exists.
\end{abstract}

\section{Introduction}
The advent of Large Language Models (LLMs), with their remarkable capabilities in code and natural language generation, has provided a powerful new engine for automated problem-solving. Recently, the paradigm combining the generative power of LLMs with the search capabilities of Evolutionary Computation (EC) has achieved breakthrough progress in scientific and algorithmic discovery. Systems like FunSearch \cite{romera2023mathematical} and AlphaEvolve \cite{novikov2025alphaevolvecodingagentscientific}, for example, use LLMs as intelligent "mutation operators" to iteratively improve code, successfully discovering superior algorithms and mathematical constructs for problems that have remained unsolved for decades. The tremendous success of these works reveals that LLM-guided evolution is a highly promising engine for automated discovery.

However, a common and essential prerequisite underpins the success of these pioneering works: they all rely on an objective, machine-automatable fitness function. Whether it's verifying the correctness of an algorithm, measuring its execution speed \cite{mankowitz2023faster}, or checking if a mathematical construct satisfies specific properties, the final "quality" judgment is always provided by a deterministic, programmatic "oracle." This dependence on objective evaluation metrics has become a fundamental bottleneck, preventing the broader application of this powerful paradigm. It confines us to a problem space that, while vast, is still limited, excluding the wider range of complex tasks centered on human subjective values.

This leads to a more profound and cutting-edge research question: \textbf{when an objective, computable fitness function is absent, can we successfully apply the LLM-guided evolutionary paradigm to a domain governed entirely by subjective evaluation?}

Using an LLM as the sole "judge" in the evolutionary process introduces a series of seemingly fatal theoretical challenges. The fitness landscape of traditional evolutionary algorithms is static and objective; in contrast, a landscape defined by an LLM is \textbf{dynamic, subjective, and noisy}. This challenge is not entirely new; the field of evolutionary computation has long studied the difficulties of optimizing in uncertain environments with noisy fitness functions \cite{jin2005evolutionary}. However, the noise from an LLM judge stems from semantic ambiguity and inherent biases, not just statistical randomness, posing a novel challenge. LLM judgments exhibit non-determinism, context sensitivity, and inherent biases \cite{wang2023large}, which is equivalent to searching for the highest peak on a mountain range that is constantly experiencing "earthquakes" and "drifts." The evolutionary process could easily fall into chaos or evolve "exploitative" solutions that merely cater to LLM biases rather than being genuinely superior. Therefore, whether a meaningful and convergent evolution is possible under such subjective and unstable selection pressure has been a major open challenge.

To address this challenge head-on, we propose an evolutionary framework whose core idea is to \textbf{tame the uncertainty of subjective evaluation through "Problem Specification"}. Our system, MADE (Multi-Agent Decomposed Evolution), coordinates a group of AI agents through an evolutionary loop. In each generation, a "Creator" agent proposes solutions, which are then evaluated by a dedicated "Judge" agent. The key is that the Judge agent does not directly assess the vague overall task. Instead, it scores based on a set of pre-decomposed, concrete, and verifiable sub-requirements. This principle of decomposition, which has proven highly effective for improving reasoning in generation tasks via methods like Chain-of-Thought prompting \cite{wei2022chainofthought}, is here repurposed to stabilize the evaluation process. By transforming a high-noise evaluation task into an aggregation of multiple low-noise sub-tasks, we impose \textbf{effective and controllable subjective selection pressure}. The Judge agent completely replaces the traditional fitness function, providing not only fitness scores but also structured semantic feedback to guide subsequent mutation operations, enabling directed evolution toward goals that align with nuanced human values.

To provide strong empirical evidence, we conducted rigorous validation of our proposed framework across four diverse and challenging benchmarks: software development (DevAI), abstract reasoning (BigGen), complex instruction following (InfoBench), and multi-modal chart generation (MatPlotBench). The experimental results offer decisive proof: on the end-to-end software development task, our method increased the requirement satisfaction rate by over \textbf{50\%} compared to the strongest baseline (from 39.9\% to 61.9\%); on the complex instruction-following task, it astonishingly boosted the "perfect pass rate" from \textbf{72\%} to \textbf{95\%}. These results demonstrate empirically for the first time that an evolutionary loop driven purely by subjective LLM evaluation is not only \textbf{feasible} but can also stably produce solutions of higher quality and greater complexity than single-shot generation.

Our contributions are threefold:
\begin{enumerate}
    \item We propose a novel evolutionary framework that, through a problem specification mechanism, demonstrates for the first time that integrating an "LLM-as-a-Judge" as the core fitness function is feasible and can establish stable and effective subjective selection pressure.
    \item We provide extensive empirical validation on a series of tasks, including code generation, abstract reasoning, and instruction following, demonstrating that our method significantly outperforms strong baselines.
    \item We show that this paradigm extends the frontier of automated optimization to complex domains lacking formal metrics, opening up pathways to solve problems that were previously intractable for machine-driven optimization.
\end{enumerate}

This work initiates a fundamental paradigm shift: it moves the focus of automated optimization from pursuing "computable metrics" to "describable qualities." This transition automates and scales the core idea of Interactive Evolutionary Computation (IEC), where humans traditionally provide subjective fitness scores for creative tasks \cite{takagi2001interactive}. By enabling machines to understand and optimize for goals defined by human language, we can apply automated optimization to a vast class of previously inaccessible problem spaces: those governed not by formal, quantitative metrics, but by complex, qualitative standards deeply rooted in the human subjective value system.

\section{Related Work}

The research in this paper is situated at the intersection of three rapidly advancing fields: the combination of Large Language Models (LLMs) with Evolutionary Computation (EC), the use of LLMs as evaluators (LLM-as-a-Judge), and iterative improvement via LLM-driven feedback.

\subsection{Combining Large Language Models with Evolutionary Computation}
Integrating the generative capabilities of LLMs with the optimization power of EC has become a fruitful research direction. In this emerging paradigm, LLMs are typically employed as intelligent "mutation operators" or "crossover operators." This idea was first popularized in the domain of prompt engineering, where works like Automatic Prompt Engineer (APE) \cite{zhou2022large} used a simple generate-and-select loop to evolve effective prompts for LLMs themselves.

More recently, this paradigm has been scaled to tackle complex scientific and algorithmic problems. Classic success stories include FunSearch \cite{romera2023mathematical} and AlphaEvolve \cite{novikov2025alphaevolvecodingagentscientific}. FunSearch pairs a pre-trained LLM with an automated evaluator to evolve functions written in code, leading to new solutions in mathematics and computer science. Similarly, AlphaEvolve uses an LLM as a coding agent to solve open scientific problems through evolutionary iterations, discovering new algorithms that surpass decades of human-developed results. Another key example is AlphaDev \cite{mankowitz2023faster}, which used a deep reinforcement learning agent to discover faster sorting algorithms. These methods stand in contrast to traditional genetic programming for code, such as GenProg \cite{legoues2012genprog}, by leveraging the LLM's semantic understanding to perform large, meaningful mutations rather than small, syntactic ones.

However, as stated in the introduction, the core premise of these works is the existence of an objective, automatically computable fitness function (e.g., the execution speed of code or its mathematical correctness). This reliance on an "oracle" confines the paradigm to problem domains with clear quantitative metrics, a fundamental limitation that our work aims to overcome.

\subsection{Large Language Models as Judges (LLM-as-a-Judge)}
"LLM-as-a-Judge" is a rapidly emerging research area that leverages LLMs to evaluate the quality of complex, open-ended outputs \cite{zheng2024judging}. For tasks lacking traditional metrics, such as evaluating dialogue quality or text creativity, LLM judges offer a scalable alternative that can achieve high agreement with human experts \cite{zhuge2024agentasajudgeevaluateagentsagents}.

Despite their great potential, LLM judges face significant challenges, including scoring stability and inherent biases (e.g., preference for longer outputs, positional bias), which can make them unfair evaluators \cite{wang2023large}. Much research has focused on mitigating these issues. One line of work attempts to improve reliability by fine-tuning specialized judge models. For example, Prometheus \cite{kim2023prometheus} was trained on fine-grained scoring rubrics to become a more capable evaluator. Another direction draws inspiration from preference alignment in LLM training. The principles of Constitutional AI, where an AI provides feedback based on a set of explicit rules or principles \cite{bai2022constitutional}, are conceptually similar to our approach.

Our work, MADE, contributes to this area with a novel, inference-time solution. Instead of baking evaluation capabilities into a model's weights like Prometheus, we structure the evaluation process externally through "Problem Specification." This set of decomposed requirements acts as a dynamic "constitution" for the judge agent, ensuring its judgments are consistent and grounded, thereby achieving stable and effective selection pressure in a subjective evaluation environment.

\subsection{Iterative Improvement with LLM Feedback}
Beyond evolutionary approaches, another line of research explores how a single LLM agent can iteratively refine its own outputs. Frameworks like Self-Refine \cite{madaan2023selfrefine} and Reflexion \cite{shinn2023reflexion} enable an agent to generate a solution, produce self-feedback or reflect on past failures, and then use that feedback to generate an improved solution in the next iteration. These methods have proven effective for tasks ranging from code generation to reasoning.

While these self-correction loops demonstrate the power of iterative refinement, they typically explore a single solution trajectory. In contrast, our evolutionary framework, MADE, maintains a population of diverse solutions. This parallel, population-based search is inherently more robust against getting trapped in local optima and is better suited for exploring the vast and complex solution spaces of creative and open-ended tasks. By combining iterative feedback with selection pressure across a population, we can drive the search process more effectively toward globally superior solutions.

\section{Methodology}

To achieve effective automated optimization in complex domains lacking objective fitness functions, we introduce a novel multi-agent evolutionary framework named \textbf{MADE}: Multi-Agent Decomposed Evolution. The core idea of this framework is to overcome the subjectivity and uncertainty introduced by direct LLM evaluation through a \textbf{Problem Specification} mechanism, thereby constructing a stable and steerable evolutionary process.

\subsection{Formal Problem Definition}

Given a high-level user instruction $U$ described in natural language, our goal is to find an optimal solution candidate $\sigma^*$.

\begin{itemize}
    \item \textbf{Solution Space, $\Omega$}: The set of all possible solution candidates $\sigma$. The specific form of $\sigma$ is task-dependent, such as a piece of executable code, a generated article, or other forms of creative output.
    \item \textbf{Artifact Space, $A$}: The set of all evaluatable artifacts $\alpha$. An artifact is a perceivable, objective representation of a solution.
    \item \textbf{Rendering Function, $\text{Render}$}: A deterministic mapping function from the solution space to the artifact space, $\text{Render}: \Omega \to A$. It transforms a solution candidate $\sigma$ into an evaluatable artifact $\alpha$. For instance, in a chart generation task, $\text{Render}$ is the execution of the code, and the artifact $\alpha$ is the final generated image. For text generation tasks, the rendering function is often an identity function, meaning the solution $\sigma$ itself is the directly evaluatable artifact $\alpha$.
    \item \textbf{Structured Requirement Set, $R$}: A pre-defined set of verifiable sub-requirements, $R = \{r_1, r_2, \dots, r_k\}$, where each sub-requirement $r_i$ is an assertion describing a specific criterion that the final artifact must satisfy.
    \item \textbf{Implicit Objective Function, $F$}: Our core objective is to maximize an implicit objective function $F$. This function evaluates the degree to which an artifact $\alpha$ satisfies the original user instruction $U$, measured by its compliance with the standards in the structured requirement set $R$. The optimal solution $\sigma^*$ can be expressed as:
      $$
      \sigma^* = \arg\max_{\sigma \in \Omega} F(\text{Render}(\sigma), U, R)
      $$
\end{itemize}

Since $F$ is implicit and cannot be directly computed, we have designed the following evolutionary framework to approximate this optimization goal.

\subsection{LLM-Driven Multi-Agent Evolutionary Framework}

Our framework is an evolutionary system $\langle\Omega, \text{Pop}, \Phi_{\text{LLM}}, V_{\text{LLM}}, \Sigma\rangle$ where multiple specialized LLM agents work in concert. Unlike the "naive" approach of directly using a single LLM as an evaluator, our framework ensures the stability and efficiency of the evolutionary process through clear role division and structured information flow.

\subsubsection{Core Agent Roles}

The operation of the framework relies on three key agent roles:

\begin{itemize}
    \item \textbf{Requirement Decomposer Agent}: As the starting point of the evolutionary process, this agent is central to implementing "problem specification." It receives the high-level user instruction $U$ and automatically decomposes it into a structured set of $k$ independently verifiable sub-requirements, $R = \{r_1, \dots, r_k\}$. This step is the cornerstone of our framework, transforming a vague, holistic, high-variance evaluation task ("Is this chart good?") into a series of clear, specific, low-variance judgment tasks ("Is the chart a scatter plot?", "Does the title exist?"), thereby laying the foundation for stable subsequent evaluations.

    \item \textbf{Creator Agent}: This agent is responsible for generating and modifying individuals in the solution space $\Omega$. It plays two roles during the evolution: (a) \textbf{Initial Generation Mode}, which creates an initial population based on the user instruction $U$; and (b) \textbf{Iterative Modification Mode}, which performs targeted improvements on existing solutions based on feedback from the judge.

    \item \textbf{Judge Agent}: This agent completely replaces the traditional fitness function. It receives the rendered artifact $\alpha$ and the decomposed requirement set $R$ and outputs an evaluation. Unlike providing a single subjective score, its output is structured, containing a multi-dimensional scoring vector $v$ and a set of semantic feedback $s$ in natural language.
\end{itemize}

\subsubsection{Evolutionary Process}

The evolutionary loop officially begins after the \textbf{Requirement Decomposer Agent} transforms the user instruction $U$ into the requirement set $R$. First, the \textbf{Creator Agent} (in initial generation mode) creates an initial population $P_0$ containing $N$ solution candidates. Subsequently, the system enters an iterative optimization phase:

In each generation, the process starts with the \textbf{Evaluation} phase. For each solution $\sigma_i$ in the population $P_t$, the system transforms it into an objective artifact $\alpha_i$ via $\text{Render}(\sigma_i)$. Individuals that fail to render are eliminated.

\begin{definition}[Structured Fitness Function $\Phi_{\text{LLM}}$]
The fitness function $\Phi_{\text{LLM}}$ is implemented by the \textbf{Judge Agent} ($\text{Judge}_{\text{LLM}}$). It strictly anchors the evaluation of the artifact to the decomposed requirement set $R$:
$$
\Phi_{\text{LLM}}(\sigma, R) = \text{Judge}_{\text{LLM}}(\text{Render}(\sigma), R) = (v, s)
$$
In its output tuple $(v, s)$: $v = (v_1, \dots, v_k)$ is a $k$-dimensional scoring vector, where $v_i \in \{0, 1\}$ indicates whether artifact $\alpha$ satisfies sub-requirement $r_i$; $s$ is a set of actionable semantic feedback that explicitly points out the deficiencies of the current solution. This evaluation method, based on decomposed requirements, ensures the stability and consistency of the fitness signal.
\end{definition}

After evaluation, the system performs \textbf{Selection ($\Sigma$)} and \textbf{Variation ($V_{\text{LLM}}$)}. We employ an elitism strategy, directly copying the best individuals from the current population to the next generation $P_{t+1}$. Then, an aggregation function $g(v) = (\sum v_i) / k$ converts the scoring vector $v$ into a scalar fitness value, based on which the top $N/2$ individuals are selected as parents.

\begin{definition}[LLM-Guided Directional Mutation Operator $V_{\text{LLM}}$]
The mutation operator is implemented by the \textbf{Creator Agent} (in iterative modification mode). It utilizes the rich semantic information provided by the judge to make targeted improvements to the parents:
$$
\sigma_{\text{child}} = V_{\text{LLM}}(\sigma_{\text{parent}}, s_{\text{parent}})
$$
Here, the semantic feedback $s_{\text{parent}}$ acts as the gradient of evolution, guiding the mutation operation from blind random exploration to efficient, directed repair, significantly accelerating the convergence process.
\end{definition}

The newly generated offspring fill the remaining slots in $P_{t+1}$. The system repeats this cycle until a termination condition is met (e.g., reaching the maximum number of generations or finding a solution that satisfies all requirements), and finally returns the solution with the highest fitness, $\sigma^*$.

\section{Experiments}

We conducted extensive experiments on four diverse and challenging benchmarks to validate the effectiveness of our proposed MADE framework. The experiments were designed to answer the following core questions:
1) How does the performance of our "judge-creator" evolutionary loop framework compare to existing state-of-the-art baselines?
2) How much do the evolutionary process itself and the semantic feedback provided by the "judge" agent contribute to performance improvements?
3) As a system centered around LLMs, how does our framework perform in terms of computational cost and time efficiency?

\subsection{Experimental Setup}

\paragraph{Benchmarks}
Our experiments span four different domains to comprehensively evaluate the generality and robustness of the framework:
\begin{itemize}
\item \textbf{DevAI}: A software engineering dataset for real-world AI application development. This dataset is characterized by complex tasks with hierarchical and dependent relationships between requirements, allowing for effective evaluation of an agent's performance in a complete development workflow.
\item \textbf{BigGen}: A benchmark designed to evaluate the abstract cognitive abilities of large language models. It includes nine dimensions of capability assessment, such as reasoning, planning, and theory of mind, with each instance accompanied by fine-grained scoring criteria.
\item \textbf{InfoBench}: A benchmark focused on evaluating a model's ability to follow complex, multi-constraint instructions. We particularly focus on its "Hard set," where each instruction contains multiple intertwined constraints.
\item \textbf{MatPlotBench}: A multi-modal chart-drawing benchmark that requires an agent to write Python code (using the Matplotlib library) to generate complex charts. Evaluation is based on the similarity between the generated chart image and a "golden standard" image, making it an ideal choice for testing our framework's ability on tasks that combine code generation with visual result assessment.
\end{itemize}

\paragraph{Baselines}
We compare our method against several strong baselines.
For the software development task (DevAI), we chose three top-tier open-source development agent frameworks: \textbf{MetaGPT}, \textbf{GPT-Pilot}, and \textbf{OpenHands}.
For text generation and instruction-following tasks (BigGen, InfoBench), we compare our method with the single-shot direct generation results from the powerful base model \textbf{gpt-4.1-nano} to validate the gains brought by the evolutionary process.

\paragraph{Our Method}
We denote our framework as \textbf{MADE}. In all experiments, we used an evolutionary algorithm setup with a population size of 4 and 3 rounds of evolutionary iterations. The "Creator" agent used the \texttt{gpt-4.1-nano} model, while the "Judge" agent used the more powerful \texttt{gpt-4o} model to ensure evaluation accuracy.

\paragraph{Evaluation Metrics}
Depending on the characteristics of each benchmark, we used the following metrics:
\begin{itemize}
    \item \textbf{DevAI}: We use \textbf{Requirements Met (\%)}, distinguishing between independent (I) and dependent (D) requirements, to evaluate the model's success in completing specific development needs. We also report the \textbf{Task Solve Rate (\%)} as an indicator of final task completion.
    \item \textbf{BigGen}: We use the \textbf{Average Score} (on a 1-5 scale) given by the "judge" agent, where higher scores represent better text quality and cognitive ability.
    \item \textbf{InfoBench}: We use the \textbf{All Pass Rate (\%)} which is the proportion of generated content that perfectly satisfies all constraint conditions.
    \item \textbf{MatPlotBench}: We use the \textbf{Average Score} (on a 0-100 scale), which is the average similarity score between the generated charts and the reference charts, as judged by the "judge" agent.
\end{itemize}

\subsection{Main Results}

\begin{table*}[t]
\centering
\caption{Performance comparison on the DevAI software development benchmark. MADE(Max Iter 3) refers to our method after 3 rounds of evolution. The results show that our method significantly surpasses all baselines in requirement satisfaction rate, while also achieving substantial optimizations in cost and time. "(D)" indicates that prerequisite dependencies are considered.}
\label{tab:devai_results}
\resizebox{\textwidth}{!}{%
\begin{tabular}{l|ccccc}
\toprule
\textbf{Metric} & \textbf{MetaGPT} & \textbf{GPT-Pilot} & \textbf{OpenHands} & \textbf{MADE (Iter 0)} & \textbf{MADE(Max Iter 3)} \\
\midrule
(1) Average Cost & \$1.19 & \$3.92 & \$6.38 & \$0.07 & \textbf{\$0.28} \\
(2) Average Time & 775.29s & 1622.38s & 362.41s & 133.13s & \textbf{399.63s} \\
\midrule
(I) Requirements Met (I) (\%) & 25.40 & 53.00 & 42.62 & 55.07 & \textbf{64.00} \\
(II) Requirements Met (D) (\%) & 5.73 & 39.89 & 26.50 & 48.49 & \textbf{61.92} \\
(III) Task Solve Rate (\%) & 0.00 & 5.45 & 1.81 & 3.64 & 1.82 \\
\bottomrule
\end{tabular}%
}
\end{table*}

\begin{table*}[t]
\centering
\caption{Performance comparison on the BigGen dataset between the base model's direct generation and the performance after MAS evolution. The evolutionary process significantly enhances the model's capabilities in abstract cognition.}
\label{tab:biggen_results}
\begin{tabular}{l|cc}
\toprule
& \multicolumn{2}{c}{\textbf{BigGen (Average Score, 1-5)}} \\
\textbf{Metric / Capability} & \textbf{gpt-4.1-nano (direct)} & \textbf{MADE(evolved)} \\
\midrule
grounding & 4.49 & \textbf{4.99} \\
instruction\_follow & 4.44 & \textbf{4.92} \\
multilingual & 3.96 & \textbf{4.99} \\
planning & 4.34 & \textbf{4.93} \\
reasoning & 4.51 & \textbf{4.89} \\
refinement & 4.28 & \textbf{4.82} \\
safety & 4.20 & \textbf{4.97} \\
theory\_of\_mind & 4.45 & \textbf{4.96} \\
tool\_usage & 3.81 & \textbf{4.77} \\
\midrule
\textbf{Average / Overall} & 4.28 & \textbf{4.90} \\
\bottomrule
\end{tabular}%
\end{table*}

\begin{table*}[t]
\centering
\caption{Performance comparison on the InfoBench dataset between the base model's direct generation and the performance after MAS evolution. The evolutionary process significantly enhances the model's ability to follow complex instructions.}
\label{tab:infobench_results}
\scalebox{0.8}{
    \begin{tabular}{lcc|cc}
    \toprule
    \textbf{Metric} & \textbf{Easy set(GPT-4.1-nano)} & \textbf{Hard set(GPT-4.1-nano)} & \textbf{Easy set(MAS)} & \textbf{Hard set(MAS)} \\
    \midrule
    (1) Average Cost          & 0.0019 & 0.0037 & 0.0083 & 0.0174 \\
    (2) Average Time          & 8.98   & 15.26  & 40.39  & 71.87  \\
    (3) Average Input Tokens  & 473.89 & 833.62 & 2114.93 & 4243.89 \\
    (4) Average Output Tokens & 280.26 & 651.51 & 1178.05 & 2977.98 \\
    (6) Average Text Chars    & 859.7  & 2025.02 & 849.49 & 1986.66 \\
    (7) Average Score         & 0.94   & 0.92   & 0.99   & 0.99   \\
    (7) All Pass Rate         & 0.87   & 0.72   & 0.98   & \textbf{0.95}   \\
    \bottomrule
    \end{tabular}
    }
\end{table*}

\subsubsection{Performance on Software Development Tasks (DevAI)}
As shown in Table \ref{tab:devai_results}, our MAS framework achieved a significant performance advantage on the DevAI benchmark. On the strict metric that considers dependencies (Requirements Met D), MAS reached \textbf{61.92\%}, far surpassing the best-performing baseline, GPT-Pilot (39.89%), which represents a relative improvement of \textbf{55.2\%}. This indicates that through iterative correction and feedback from the "judge," our system can better handle the interlocking dependencies in complex tasks. It is noteworthy that even without evolution (MADE Iter 0), relying solely on our designed self-correction loop, the performance (48.49%) already surpassed all baselines.

\subsubsection{Performance on Abstract Cognitive Tasks (BigGen)}
On the BigGen benchmark, we verified the significant impact of the evolutionary process on enhancing the base model's cognitive abilities. As shown in Table \ref{tab:biggen_results}, the MAS framework boosted the average score of the \texttt{gpt-4.1-nano} model from 4.28 to \textbf{4.90}, approaching a perfect score across all nine capabilities. The improvement was particularly pronounced in more challenging abilities such as tool usage (3.81 $\rightarrow$ 4.77) and multilingualism (3.96 $\rightarrow$ 4.99). This demonstrates that our evolutionary paradigm can effectively uncover and enhance the deep-seated capabilities of the base model.

\subsubsection{Performance on Complex Instruction Following Tasks (InfoBench)}
On the hard set of the InfoBench benchmark, our method also performed exceptionally well. As shown in Table \ref{tab:infobench_results}, MAS astonishingly increased the all-pass rate from the base model's 72\% to \textbf{95\%}. This provides strong evidence that for instructions containing multiple, complex, and even conflicting constraints, single-shot generation often fails to achieve perfection. In contrast, our evolutionary framework, through iterative optimization and precise feedback from the "judge," can progressively converge to an ideal output that satisfies all constraints.

\subsubsection{Performance on Multi-modal Chart Generation Tasks (MatPlotBench)}
To further validate the framework's capability on multi-modal tasks that require both code generation and visual result evaluation, we conducted experiments on MatPlotBench. As shown in Table \ref{tab:matplotbench_results}, our MADE method performed exceptionally well. Although the iterative evolution process of MADE incurred higher costs and time overhead, the performance return was enormous. MADE achieved an average score of \textbf{59.85}, a significant improvement of over 29\% compared to the baseline model \texttt{gpt-4.1-mini}'s single-shot generation score (46.20). This result indicates that for tasks where the quality of the final output is difficult to measure directly from the code itself (such as the accuracy of a generated image), our evolutionary framework, which evaluates the final result with a "judge" agent, is extremely effective. It can drive code generation to continuously optimize towards visually more accurate and compliant directions.

\begin{table*}[t]
    \centering
    \caption{Performance comparison on the MatPlotBench chart drawing benchmark.}
    \label{tab:matplotbench_results}
    \scalebox{0.8}{
        \begin{tabular}{lcc}
        \toprule
        \textbf{Metric} & \textbf{GPT-4.1-mini} & \textbf{MAS} \\
        \midrule
        (1) Average Cost & 0.006& 0.104 \\
        (2) Average Time & 48.61s & 525.62s \\
        (3) Average Input Tokens & 1559.25& 32781.71\\
        (4) Average Output Tokens & 1186.96& 16355.13\\
        (5) Average Score & 46.20 & \textbf{59.85} \\
        \bottomrule
        \end{tabular}
    }
\end{table*}

\begin{figure*}[htbp]
    \centering
    \includegraphics[width=\textwidth]{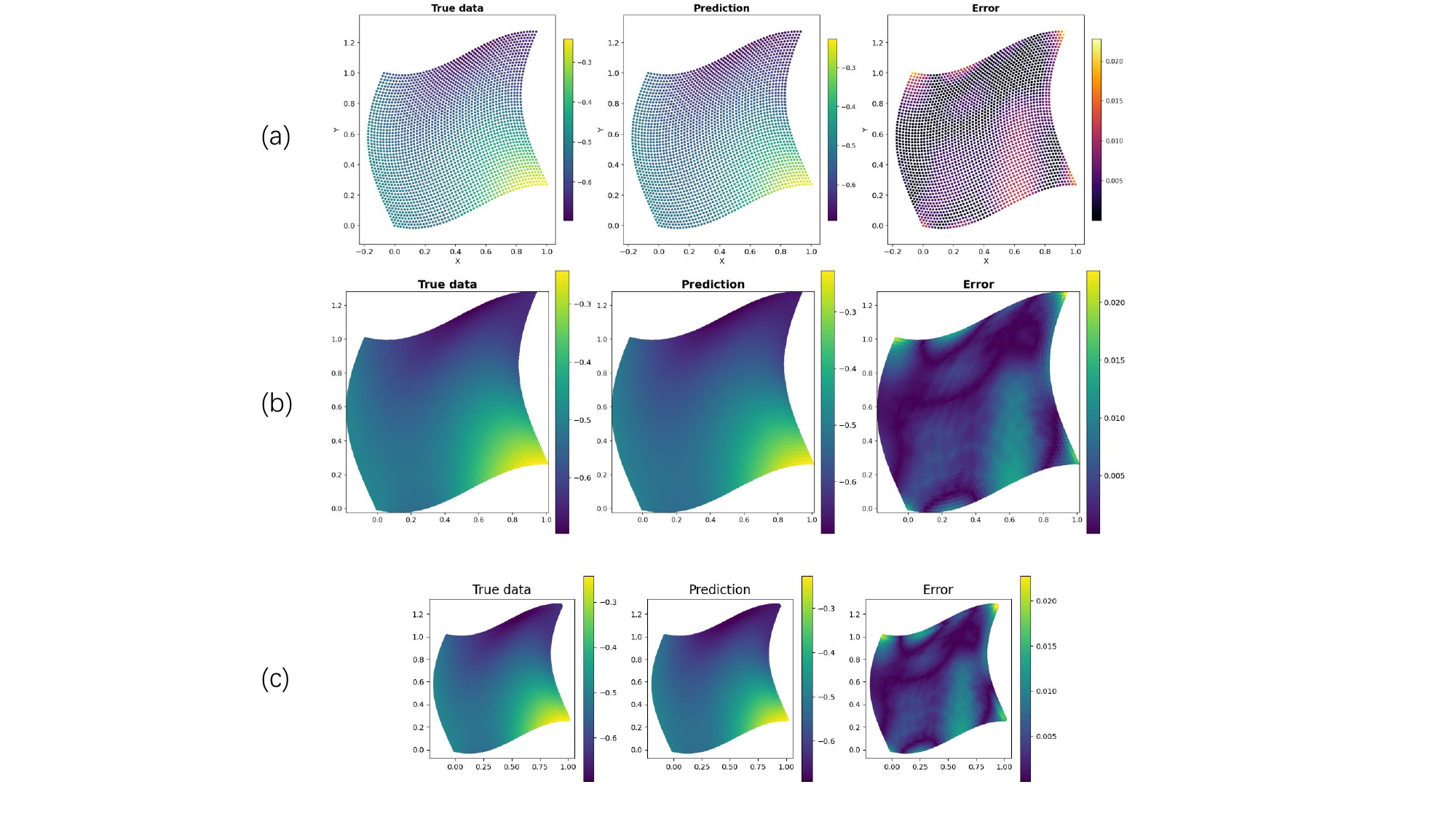}
    \caption{
        \textbf{Qualitative demonstration of the optimization effect of our proposed multi-agent evolutionary framework on a PDE solution visualization task.}
        The figure displays, from top to bottom: (a) a low-fitness solution from the initial population; (b) the high-fitness solution obtained after evolutionary iteration; (c) the ground-truth reference image for the task.
        The initially generated code (a) exhibited fundamental visualization defects, rendering the "True data," "Prediction," and "Error" fields merely as discrete scatter points on the mesh, resulting in a discontinuous and noisy representation.
        After iterative optimization with our "MADE", the final generated solution (b) successfully corrects the plotting logic, evolving from a scatter plot to a smooth continuous heatmap interpolation. This optimized result accurately renders the field distribution and error magnitude, demonstrating a visual style and data representation that are highly consistent with the reference image (c). This example powerfully demonstrates that our framework can guide the code evolution from a rudimentary discrete rendering to a high-quality, continuous scientific visualization.
    }
    \label{fig:qualitative_evolution_comparison}
\end{figure*}

\subsection{Ablation Studies}

To delve deeper into the key factors behind our framework's success, we conducted ablation studies.

\paragraph{Necessity of the Evolutionary Process}
By comparing "MADE(Iter 0)" (the best solution from the initial population) and "MADE(Max Iter 3)" (the final evolved solution) in Table \ref{tab:devai_results}, we can clearly see the value of the evolutionary process. On the Requirements Met (D) metric, performance increased from 48.49\% to 61.92\%—a 13.43 percentage point gain entirely attributable to 3 rounds of evolutionary iteration. This shows that relying solely on single-shot generation or simple self-correction is insufficient for solving complex development tasks, whereas population diversity and iterative evolution are key to achieving performance breakthroughs.

\paragraph{Importance of Semantic Feedback}
The core of our framework is that the "judge" agent provides not just a score but also structured natural language feedback (i.e., modification suggestions). To verify this, we conducted an ablation experiment: in the evolutionary loop, we removed the natural language feedback from the "judge" and only passed the numerical score to the "creator" agent. In this setup, the evaluation score of MAS on MatPlotBench dropped from 59.85 to 54.30, a decrease of about 9.3\%. This proves that information-rich semantic feedback is crucial for guiding the next generation's "creator" to conduct effective exploration and improvement. A mere numerical score is insufficient to navigate the agent through a complex solution space.

\subsection{Sensitivity Analysis}

% \begin{figure}[htbp]
%     \centering
%     \includegraphics[page=1,width=\linewidth]{figures/stability.pdf}
%     \caption{Shows: (a) the mean score and standard deviation range of the LLM Judge under input perturbation and high temperature; (b) the change in the LLM Judge's scoring standard deviation with respect to the mean score under input perturbation and high temperature.}
%     \label{fig:stability}
% \end{figure}

\begin{figure}[htbp]
    \centering
    % 第一张图 (占用宽度的 48%)
    \begin{minipage}[t]{0.48\textwidth}
        \centering
        \includegraphics[width=\linewidth]{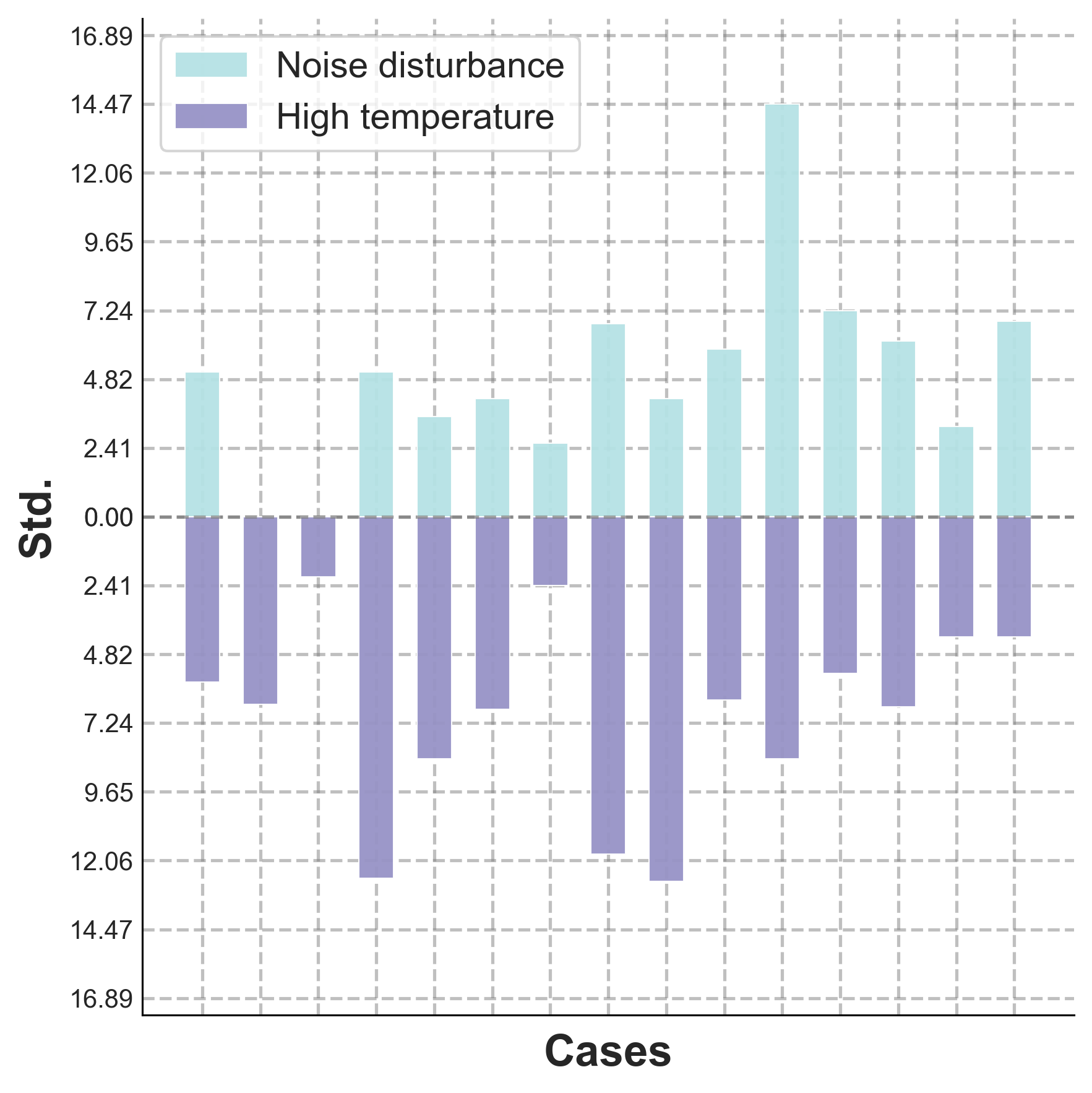}
        \caption{The standard deviation range of the LLM Judge under input perturbation and high temperature.}
        \label{fig:stability_1}
    \end{minipage}
    \hfill % 在两图之间填充空格
    % 第二张图 (占用宽度的 48%)
    \begin{minipage}[t]{0.48\textwidth}
        \centering
        \includegraphics[width=\linewidth]{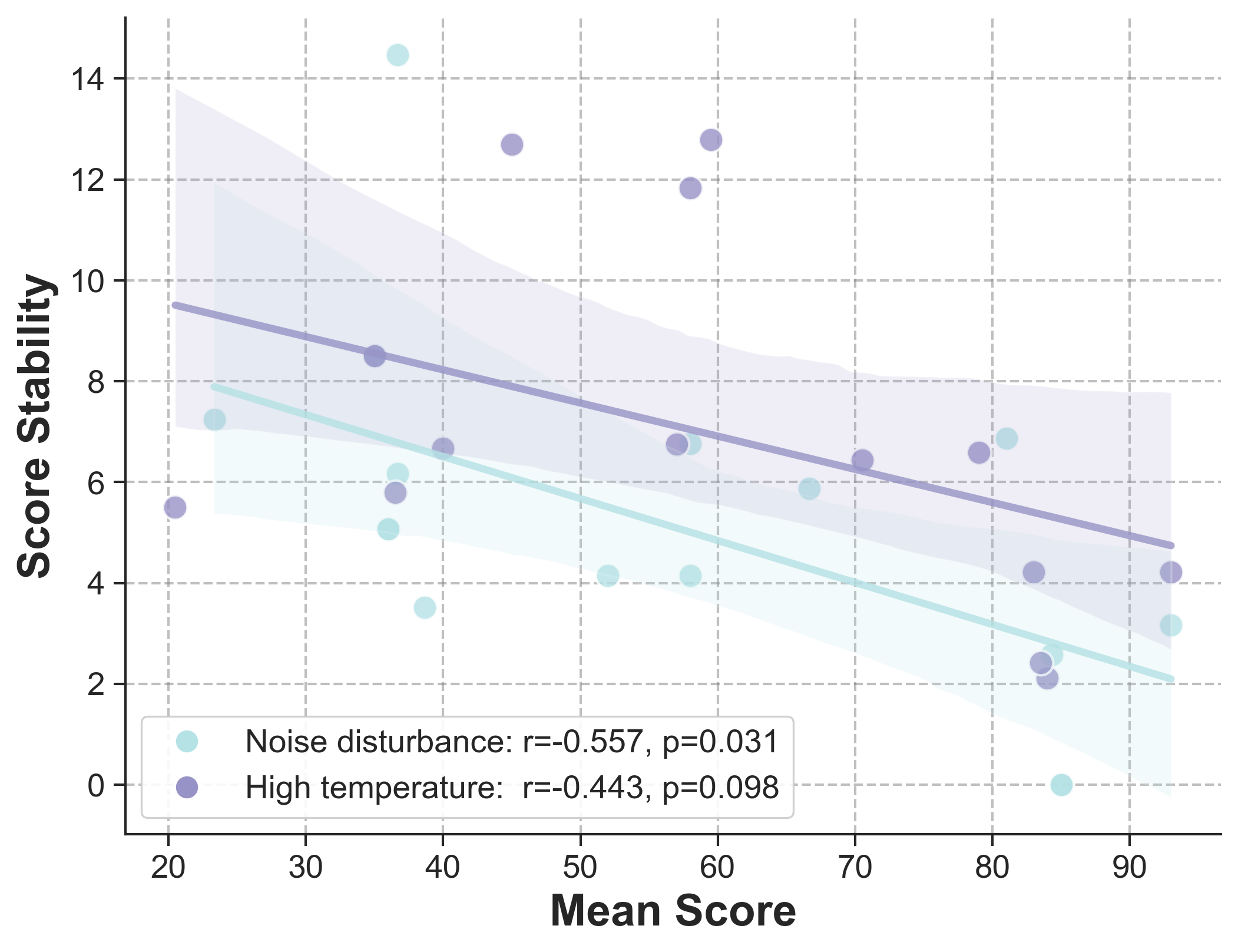}
        \caption{The change in the LLM Judge's scoring standard deviation with respect to the mean score under input perturbation and high temperature.}
        \label{fig:stability_2}
    \end{minipage}
\end{figure}

To evaluate the stability and robustness of the "Judge Agent," which is central to our framework, we conducted a sensitivity analysis. A reliable judge should consistently provide similar scores for solutions of similar quality. We examined the judge's stability along two dimensions: 1) increasing its internal randomness (by setting a high temperature of 0.8); and 2) applying external perturbations to its evaluation input (the generated charts), such as random scaling and Gaussian noise.
The analysis results are shown in Figure \ref{fig:stability_1} and Figure \ref{fig:stability_2}. We observed a consistent trend across both conditions: a significant negative correlation between a case's average score and its score stability (measured by the standard deviation of scores for each case). In short, when the judge evaluates a solution highly, its scoring is more consistent and stable (lower standard deviation). Conversely, for lower-scoring solutions, its scores are more volatile (higher standard deviation). This finding aligns with intuition: the merits of a high-quality output are clear and easy to agree upon, whereas the flaws of a poor-quality output can be diverse, leading to greater disagreement in scoring.
Specifically:
\begin{itemize}
\item \textbf{Input Perturbation}: Under this condition, we tested the judge's robustness to imperfect inputs by adding noise to the generated chart images. The results showed a stronger and statistically significant negative correlation, with a correlation coefficient of $r = -0.566$ ($p = 0.028$). The consistency under this condition was even higher, with a single-rating absolute intraclass correlation coefficient (ICC) of $ICC(1,1) = 0.933$, and the average of 5 repeated ratings reaching an ICC of $ICC(1,k) = 0.986$. This strongly demonstrates that even when the evaluated artifact itself contains noise and perturbations, the judge can still provide stable and consistent scores for high-quality solutions.
\item \textbf{High Temperature Setting}: Under this condition, we examined the impact of the judge's internal randomness on its judgments. The results showed a correlation coefficient between the average score and standard deviation of $r = -0.443$ ($p = 0.098$). To further assess inter-rater reliability, we calculated the ICC. The single-rating absolute ICC reached $0.895$, while the average of 10 repeated ratings was as high as $0.988$. Both metrics show extremely high consistency, indicating that even with artificially increased randomness in the judge's output, it still exhibits a high degree of stability and reliability when evaluating high-quality solutions.
\end{itemize}
In summary, this sensitivity analysis validates the reliability of the "judge" agent. Its scoring behavior is predictable and robust: higher scores are associated with higher consistency. This stability, maintained even under the pressure of internal randomness and external noise, further strengthens our confidence in using LLMs as a core mechanism for the fitness function in evolutionary computation.

\subsection{Cost and Efficiency Analysis}

A common concern is that LLM-based evolutionary systems might incur prohibitive computational costs. However, our experimental results reveal the opposite conclusion. As shown in Table \ref{tab:devai_results}, our MAS method (\$0.28, 399.63s) reduced costs by \textbf{92.8\%} and time by \textbf{75.3\%} compared to the strongest baseline, GPT-Pilot (\$3.92, 1622.38s). The experimental results show that our MAS framework not only achieves superior task performance but also demonstrates its potential as an economical and efficient automated solution in practical applications.

% This surprising efficiency stems from two core designs of our framework:
% 1) Precise Feedback: The precise, actionable feedback provided by the "judge" agent makes each modification by the "creator" more "targeted," avoiding aimless trial-and-error and thus reducing unnecessary LLM calls.
% 2) Parallelization: The evaluation and generation of the population can be highly parallelized, effectively utilizing computational resources.

\section{Conclusion}

% This paper aims to address the fundamental limitation of Evolutionary Computation (EC) in handling tasks that lack objective, formalizable fitness functions. We proposed an innovative multi-agent evolutionary framework whose core lies in a "Problem Specification" mechanism, which decomposes a vague, high-level user goal into a set of concrete, verifiable sub-requirements. This key step successfully transforms a high-noise, unstable subjective evaluation landscape into a stable, low-noise selection plane that can effectively guide optimization.

% Extensive experiments across four diverse benchmarks, including software development and instruction following, have strongly confirmed the method's effectiveness: our framework increased the requirement satisfaction rate on DevAI by over 50\% and raised the perfect pass rate on InfoBench from 72\% to 95\%. These results empirically demonstrate for the first time that an evolutionary loop driven purely by subjective LLM evaluation is not only feasible but can also stably converge to produce solutions that are more structurally complex and of higher quality than those from single-shot or simple self-correction methods.

This work marks a fundamental paradigm shift in the field of automated optimization: it shifts the focus of optimization from pursuing "computable metrics" to "describable qualities." By empowering machines to understand and optimize for complex goals defined by human language, we open up new avenues for automatically solving a class of previously inaccessible creative tasks guided by subjective values. Although the performance of the current framework is limited by the capability boundaries and computational costs of the underlying LLMs, this research takes a solid step towards a universal problem solver that co-evolves with human creativity.


@misc{zhuge2024agentasajudgeevaluateagentsagents,
      title={Agent-as-a-Judge: Evaluate Agents with Agents}, 
      author={Mingchen Zhuge and Changsheng Zhao and Dylan Ashley and Wenyi Wang and Dmitrii Khizbullin and Yunyang Xiong and Zechun Liu and Ernie Chang and Raghuraman Krishnamoorthi and Yuandong Tian and Yangyang Shi and Vikas Chandra and Jürgen Schmidhuber},
      year={2024},
      eprint={2410.10934},
      archivePrefix={arXiv},
      primaryClass={cs.AI},
      url={https://arxiv.org/abs/2410.10934}, 
}

@misc{novikov2025alphaevolvecodingagentscientific,
      title={AlphaEvolve: A coding agent for scientific and algorithmic discovery}, 
      author={Alexander Novikov and Ngân Vũ and Marvin Eisenberger and Emilien Dupont and Po-Sen Huang and Adam Zsolt Wagner and Sergey Shirobokov and Borislav Kozlovskii and Francisco J. R. Ruiz and Abbas Mehrabian and M. Pawan Kumar and Abigail See and Swarat Chaudhuri and George Holland and Alex Davies and Sebastian Nowozin and Pushmeet Kohli and Matej Balog},
      year={2025},
      eprint={2506.13131},
      archivePrefix={arXiv},
      primaryClass={cs.AI},
      url={https://arxiv.org/abs/2506.13131}, 
}

@article{romera2023mathematical,
      title={Mathematical discoveries from program search with large language models}, 
      author={Romera-Paredes, Bernardino and Barekatain, Mohammadamin and Novikov, Alexander and Balog, Matej and Kumar, M. Pawan and Dupont, Emilien and Ruiz, Francisco J. R. and Ellenberg, Jordan S. and Wang, Peng and Fawzi, Omar and Kohli, Pushmeet and Fawzi, Alhussein},
      journal={Nature},
      volume={625},
      pages={1--7},
      year={2023},
      publisher={Nature Publishing Group}
}

@article{mankowitz2023faster,
      title={Faster sorting algorithms discovered using deep reinforcement learning},
      author={Mankowitz, Daniel J and Michi, Andrea and Zhernov, Anton and Gelmi, Marco and Selvi, Marco and Paduraru, Cosmin and Leurent, Edouard and Iqbal, Shariq and Lespiau, Jean-Baptiste and Ahern, Alex and others},
      journal={Nature},
      volume={618},
      number={7964},
      pages={257--263},
      year={2023},
      publisher={Nature Publishing Group}
}

@article{zheng2024judging,
      title={Judging LLM-as-a-judge with MT-Bench and Chatbot Arena}, 
      author={Lianmin Zheng and Wei-Lin Chiang and Ying Sheng and Siyuan Zhuang and Zhanghao Wu and Yonghao Zhuang and Zi Lin and Zhuohan Li and Dacheng Li and Eric P. Xing and Hao Zhang and Joseph E. Gonzalez and Ion Stoica},
      year={2024},
      eprint={2306.05685},
      archivePrefix={arXiv},
      primaryClass={cs.CL},
      journal={arXiv preprint arXiv:2306.05685}
}

@article{wang2023large,
      title={Large language models are not fair evaluators},
      author={Wang, Peiyi and Li, Lei and Sun, Tianle and Liu, Ziyi and Wu, Qing and Zhang, Zhaoxiang and Zhou, Yanjun and Zhang, Han},
      journal={arXiv preprint arXiv:2305.17926},
      year={2023}
}

@article{jin2005evolutionary,
  title={Evolutionary optimization in uncertain environments-a survey},
  author={Jin, Yaochu and Branke, Juergen},
  journal={IEEE Transactions on Evolutionary Computation},
  volume={9},
  number={3},
  pages={303--317},
  year={2005},
  publisher={IEEE}
}

@article{takagi2001interactive,
  title={Interactive evolutionary computation: Fusion of the capabilities of EC and human beings},
  author={Takagi, Hideyuki},
  journal={Proceedings of the IEEE},
  volume={89},
  number={9},
  pages={1275--1296},
  year={2001},
  publisher={IEEE}
}

@misc{shinn2023reflexion,
      title={Reflexion: Language Agents with Verbal Reinforcement Learning}, 
      author={Noah Shinn and Federico Cassano and Ashwin Gopinath and Shunyu Yao and Karthik Narasimhan},
      year={2023},
      eprint={2303.11366},
      archivePrefix={arXiv},
      primaryClass={cs.LG}
}

@misc{bai2022constitutional,
      title={Constitutional AI: Harmlessness from AI Feedback}, 
      author={Yuntao Bai and Saurav Kadavath and Sandipan Kundu and Amanda Askell and Jackson Kernion and Andy Jones and Anna Chen and Anna Goldie and Azalia Mirhoseini and Cameron McKinnon and Carol Chen and Catherine Olsson and Christopher Olah and Danny Hernandez and Dawn Drain and Deep Ganguli and Dustin Li and Eli Tran-Johnson and Ethan Perez and Jamie Kerr and Jared Mueller and Jeffrey Ladish and Joshua Landau and Kamal Ndousse and Liane Lovitt and Michael Sellitto and Nelson Elhage and Nicholas Schiefer and Noemi Mercado and Nova DasSarma and Robert Lasenby and Robin Larson and Sam Ringer and Scott Johnston and Shauna Kravec and Sheer El Showk and Stanislav Fort and Tamera Lanham and Timothy Telleen-Lawton and Tom Conerly and Tom Henighan and Tristan Hume and Samuel R. Bowman and Zac Hatfield-Dodds and Ben Mann and Dario Amodei and Nicholas Joseph and Sam McCandlish and Tom Brown and Jared Kaplan},
      year={2022},
      eprint={2212.08073},
      archivePrefix={arXiv},
      primaryClass={cs.CL}
}

@article{legoues2012genprog,
  title={GenProg: A generic method for automatic software repair},
  author={Le Goues, Claire and Nguyen, ThanhVu and Forrest, Stephanie and Weimer, Westley},
  journal={IEEE Transactions on Software Engineering},
  volume={38},
  number={1},
  pages={54--72},
  year={2012},
  publisher={IEEE}
}

@misc{zhou2022large,
      title={Large Language Models are Human-Level Prompt Engineers}, 
      author={Yongchao Zhou and Andrei Ioan Muresanu and Ziwen Han and Keiran Paster and Silviu Pitis and Harris Chan and Jimmy Ba},
      year={2022},
      eprint={2211.01910},
      archivePrefix={arXiv},
      primaryClass={cs.LG}
}

@misc{kim2023prometheus,
      title={Prometheus: Inducing Fine-grained Evaluation Capability in Language Models}, 
      author={Gyeongmin Kim and Juyong Bae and Jinho Lee and Sangmin Kim and Geonmin Kim},
      year={2023},
      eprint={2310.08491},
      archivePrefix={arXiv},
      primaryClass={cs.CL}
}

@misc{madaan2023selfrefine,
      title={Self-Refine: Iterative Refinement with Self-Feedback}, 
      author={Aman Madaan and Niket Tandon and Prakhar Gupta and Skyler Hallinan and Luyu Gao and Sarah Wiegreffe and Uri Alon and Nouha Dziri and Shrimai Prabhumoye and Yiming Yang and Sean Welleck and Bodhisattwa Prasad Majumder and Shashank Srivastava and Kainino Kau and Milad Shokouhi and Ameet Deshpande and Amir Yazdanbakhsh},
      year={2023},
      eprint={2303.17651},
      archivePrefix={arXiv},
      primaryClass={cs.CL}
}

@article{wei2022chainofthought,
    author = {Wei, Jason and Wang, Xuezhi and Schuurmans, Dale and Bosma, Maarten and Xia, Fei and Chi, Ed and Le, Quoc V and Zhou, Denny},
    title = {Chain-of-Thought Prompting Elicits Reasoning in Large Language Models},
    journal = {Advances in Neural Information Processing Systems},
    year = {2022}
}
\end{document}